\documentclass[12pt]{article}

\usepackage{epsf}
\usepackage{epsfig}
\usepackage{cite}
\usepackage{amsmath}
\usepackage{graphics}

\begin{document}

\setlength{\textheight}{21.5cm}
\setlength{\oddsidemargin}{0.cm}
\setlength{\evensidemargin}{0.cm}
\setlength{\topmargin}{0.cm}
\setlength{\footskip}{1cm}
\setlength{\arraycolsep}{2pt}

\renewcommand{\thefootnote}{\#\arabic{footnote}}
\setcounter{footnote}{0}

\newcommand{\gtrsim}{ \mathop{}_{\textstyle \sim}^{\textstyle >} }
\newcommand{\lesssim}{ \mathop{}_{\textstyle \sim}^{\textstyle <} }
\newcommand{\rem}[1]{{\bf #1}}
\renewcommand{\thefootnote}{\fnsymbol{footnote}}
\setcounter{footnote}{0}
\def\thefootnote{\fnsymbol{footnote}}

\hfill July 2009\\

\vskip .5in

\begin{center}

\bigskip
\bigskip

{\Large \bf Analysis of Quark Mixing Using Binary Tetrahedral
Flavor Symmetry}

\vskip .45in

{\bf David A. Eby\footnote{daeby@physics.unc.edu}, Paul H. Frampton\footnote{frampton@physics.unc.edu}
and Shinya Matsuzaki\footnote{synya@physics.unc.edu}} 

\vskip .3in

{\it Department of Physics and Astronomy, University of North Carolina,
Chapel Hill, NC 27599-3255.}

\end{center}

\vskip .4in 
\begin{abstract}
Using the binary tetrahedral group $T^{'}$,
the three angles and phase of the quark CKM
mixing matrix are pursued by symmetry-breaking
which involves $T^{'}$-doublet VEVs and
the Chen-Mahanthappa CP-violation mechanism.
The NMRT$^{'}$M, Next-to-Minimal-Renormalizable
-T$^{'}$-Model is described, and its one
parameter comparison to experimental data
is explored.
\end{abstract}

\renewcommand{\thepage}{\arabic{page}}
\setcounter{page}{1}
\renewcommand{\thefootnote}{\#\arabic{footnote}}

\newpage

\section{Introduction}

\bigskip

To go beyond the standard model based on
$G=SU(3) \times SU(2) \times U(1)$ generally
has the aim of relating some of the many 
parameters therein. Well-known
possibilities include grand unification $G \in G_{GUT}$, 
otherwise extending the gauge group $G \in G'$, supersymmetry,
technicolor, and finally horizontal or flavor symmetry $G_F$,
a global group commuting with $G$.

\bigskip

In the present paper we study further the use of $G_F$,
in particular the choice $G_F =T^{'}$, the binary tetrahedral group.
This group can combine the advantages of its central quotient
$T \equiv A_4$ for leptons with the incorporation
of three quark families in a $(2+1)$ pattern with
the third much heavier family treated asymmetrically.

\bigskip

We shall employ Higgs scalars which are all electroweak
doublets. An alternative approach would be to use
electroweak singlets, so-called {\it flavons},
but that would necessitate non-renormalizable 
or irrelevant operators which we eschew.

\bigskip

In recent work, two of the present authors, together with
Kephart\cite{FKM}, presented a simplified model
based on $T^{'}$ flavor symmetry. The principal simplification
was that the CKM mixing angles
\footnote{Note that here upper case $\Theta_{ij}$ refer to quarks (CKM)
and lower case $\theta_{ij}$ will refer to neutrinos (PMNS).}
involving the third quark
family were taken to vanish $\Theta_{23} = \Theta_{13} = 0$.

\bigskip

In terms of the scalar field content, all scalar fields
are taken to be doublets under electroweak
$SU(2)$ with vacuum values which underly the symmetry breaking.
Great simplification was originally achieved by the device
of restricting scalar fields to irreducible representations
of $T^{'}$ which are singlets and triplets only, without any $T^{'}$ doublets.
There was a good reason for this because the admission of
$T^{'}$-doublet scalars enormously complicates the symmetry
breaking.  This enabled the isolation of the Cabibbo angle $\Theta_{12}$
and to a very reasonable prediction thereof, namely
\cite{FKM} $\tan 2 \Theta_{12} = (\sqrt{2})/3$.

\bigskip

Within the same simplified model, in a subsequent paper\cite{EFM},
the departure of $\Theta_{12}$ from this
$T^{'}$ prediction was used to make predictions for the departure
of the neutrino PMNS angles $\theta_{ij}$ from their
tribimaximal values\cite{HPS}.
Also in that model \cite{FM}, we
suggested a smoking-gun $T^{'}$ prediction
for leptonic decay of the standard model Higgs scalar.
Other related works are
\cite{FMA4,Ma,Ma2,Altarelli,Kephart0,Feruglio,mahanthappa,HS,textures,marfatia,FGY}.

\bigskip

In the present article, we examine the addition
of $T^{'}$-doublet scalars. As anticipated in \cite{FKM}, this
allows more possibilities of $T^{'}$ symmetry breaking
and permits non-zero values for $\Theta_{23}$, $\Theta_{13}$
and $\delta_{KM}$.  
We present an explicit $(T^{'} \times Z_2)$
model and investigate
for all the CKM angles.

\bigskip

To understand the incorporation of $T^{'}$-doublet
scalars and to make the present article self-contained,
it is necessary to review the previous simplified
model employed in \cite{FKM,EFM,FM} in which $T^{'}$-doublet
scalars were deliberately excluded in order to isolate
the Cabibbo angle $\Theta_{12}$.
We here adopt the global symmetry $(T^{'} \times Z_2)$. 

\bigskip

Note that we focus on a renormalizable model with
few if any free parameters and focus on the mixing
matrix rather than on masses as the former is more
likely to have a geometrical interpretation while
without adding many extra parameters the masses
are unfortunately not naturally predicted. This is
especially true for the lighter quarks; for
the $t$ quark the flavor group assignments
allow it much heavier mass.

\bigskip

\section{The Previous Simplified Model, MRT$^{'}$M}

\bigskip

By $MRT^{'}M$, we mean Minimal Renormalizable $T^{'}$ Model.
Actually the global symmetry, to restrict the Yukawa couplings
is $(T^{'} \times Z_2)$.

\bigskip

Left-handed quark doublets \noindent $(t, b)_L, (c, d)_L, (u, d)_L$
are assigned under $(T^{'} \times Z_2)$ as 

\begin{equation}
\begin{array}{cc}
\left( \begin{array}{c} t \\ b \end{array} \right)_{L}
~ {\cal Q}_L ~~~~~~~~~~~ ({\bf 1_1}, +1)   \\
\left. \begin{array}{c} \left( \begin{array}{c} c \\ s \end{array} \right)_{L}
\\
\left( \begin{array}{c} u \\ d \end{array} \right)_{L}  \end{array} \right\}
Q_L ~~~~~~~~ ({\bf 2_1}, +1),
\end{array}
\label{qL}
\end{equation}

\noindent and the six right-handed quarks as

\begin{equation}
\begin{array}{c}
t_{R} ~~~~~~~~~~~~~~ ({\bf 1_1}, +1)   \\
b_{R} ~~~~~~~~~~~~~~ ({\bf 1_2}, -1)  \\
\left. \begin{array}{c} c_{R} \\ u_{R} \end{array} \right\}
{\cal C}_R ~~~~~~~~ ({\bf 2_3}, -1)\\
\left. \begin{array}{c} s_{R} \\ d_{R} \end{array} \right\}
{\cal S}_R ~~~~~~~~ ({\bf 2_2}, +1).
\end{array}
\label{qR}
\end{equation}

\bigskip

\noindent The leptons are assigned as
\begin{equation}
\begin{array}{ccc}
\left. \begin{array}{c}
\left( \begin{array}{c} \nu_{\tau} \\ \tau^- \end{array} \right)_{L} \\
\left( \begin{array}{c} \nu_{\mu} \\ \mu^- \end{array} \right)_{L} \\
\left( \begin{array}{c} \nu_e \\ e^- \end{array} \right)_{L}
\end{array} \right\}
L_L  (3, +1)  &
\begin{array}{c}
~ \tau^-_{R}~ (1_1, -1)   \\
~ \mu^-_{R} ~ (1_2, -1) \\
~ e^-_{R} ~ (1_3, -1)  \end{array}
&
\begin{array}{c}
~ N^{(1)}_{R} ~ (1_1, +1) \\
~ N^{(2)}_R ~ (1_2, +1) \\
~ N^{(3)}_{R} ~ (1_3, +1),\\  \end{array}
\end{array}
\end{equation}

\bigskip

Next we turn to the symmetry breaking and the necessary
scalar sector with its own potential
\footnote{The scalar potential will not be examined
explicitly. We assume that it has enough parameters 
to accommodate the required VEVs in a finite neighborhood of 
parameter values.}
and Yukawa coupling to the fermions, leptons and quarks.

\bigskip
\bigskip

The scalar fields in the previous simplified model 
were namely the two $T^{'}$ triplets and two $T^{'}$ singlets
\begin{equation}
H_3(3, +1); ~ H_3^{'}(3, -1); ~ H_{1_1}(1_1, +1); ~ H_{1_3}(1_3, -1)
\label{oldscalars}
\end{equation}
which leads to
CKM angles $\Theta_{23} = \Theta_{13} = 0$. That
model was used to derive a formula for
the Cabibbo angle\cite{FKM}, to predict corrections\cite{EFM}
to the tribimaximal values\cite{HPS} of PMNS neutrino
angles, and to make a prediction for
Higgs boson decay\cite{FM}.

\bigskip
\bigskip

The Yukawa couplings 
for the $T^{'}$-triplet and $T^{'}$-singlet scalars
were as follows:

\begin{eqnarray}
{\cal L}_Y
&=&
\frac{1}{2} M_1 N_R^{(1)} N_R^{(1)} + M_{23} N_R^{(2)} N_R^{(3)} \nonumber \\
& & + \Bigg\{
Y_{1} \left( L_L N_R^{(1)} H_3 \right) + Y_{2} \left(  L_L N_R^{(2)}  H_3
\right) + Y_{3}
\left( L_L N_R^{(3)} H_3 \right)  \nonumber \\
&& +
Y_\tau \left( L_L \tau_R H'_3 \right)
+ Y_\mu  \left( L_L \mu_R  H'_3 \right) +
Y_e \left( L_L e_R H'_3 \right)
\Bigg\} \nonumber \\
&& + Y_t ( \{{\cal Q}_L\}_{\bf 1_1}  \{t_R\}_{\bf 1_1} H_{\bf 1_1}) \nonumber \\
&&
+ Y_b (\{{\cal Q}_L\}_{\bf 1_1} \{b_R\}_{\bf 1_2} H_{\bf 1_3} ) \nonumber \\
&&
+ Y_{{\cal C}} ( \{ Q_L \}_{\bf 2_1} \{ {\cal C}_R \}_{\bf 2_3} H^{'}_{\bf 3})
\nonumber \\
&&
+ Y_{{\cal S}} ( \{ Q_L \}_{\bf 2_1} \{ {\cal S}_R \}_{\bf 2_2} H_{\bf 3})
\nonumber \\
&&
+ {\rm h.c.}. 
\end{eqnarray}

\section{Choice of the Present Model, NMRT$^{'}$M}

\bigskip

By $NMRT^{'}M$ we mean Next-to Minimal Renormalizable $T^{'}$ Model.

\bigskip

We introduce one $T^{'}$ doublet scalar in an explicit model. 
Non-vanishing $\Theta_{23}$ and $\Theta_{13}$ will be induced by
symmetry breaking due to the addition
the $T^{'}$ doublet scalar.

\bigskip

The possible choices under $(T^{'} \times Z_2)$ for the new scalar field are:

\begin{equation}
{\bf A} ~~~ H_{2_1}(2_1, +1)
\label{newscalarA}
\end{equation}

\begin{equation}
{\bf B} ~~~ H^{'}_{2_3}(2_3, -1)
\label{newscalarB}
\end{equation}

\begin{equation}
{\bf C} ~~~ H^{'}_{2_2}(2_2, -1)
\label{newscalarC}
\end{equation}

\begin{equation}
{\bf D} ~~~ H_{2_3}(2_3, +1)
\label{newscalarD}
\end{equation}

\bigskip

\noindent The fields in Eqs.(\ref{newscalarA},\ref{newscalarB},\ref{newscalarC},\ref{newscalarD})
allow respectively 
Yukawa couplings:

\begin{equation}
{\bf A} ~~~ Y_{Qt} Q_L t_R H_{2_1} + h.c.
\label{newYukawaA}
\end{equation}

\begin{equation}
{\bf B} ~~~ Y_{Qb} Q_L b_R H^{'}_{2_3} + h.c.
\label{newYukawaB}
\end{equation}

\begin{equation}
{\bf C} ~~~ Y_{{\cal Q}{\cal C}} {\cal Q}_L {\cal C}_R H^{'}_{2_2} + h.c.
\label{newYukawaC}
\end{equation}

\begin{equation}
{\bf D} ~~~ Y_{{\cal Q}{\cal S}} {\cal Q}_L {\cal S}_R H_{2_3} + h.c.
\label{newYukawaD}
\end{equation}

This leads potentially to different extensions of the $MRT^{'}M$. For
simplicity we keep only one additional term, {\bf D},
inspired by the Chen-Mahanthappa mechanism\cite{Chen}
for CP violation.
We shall keep $Y_{{\cal Q}{\cal S}}$ real and CP violation
will arise from the imaginary part of $T^{'}$ Clebsch-Gordan
coefficients.

\bigskip

\noindent The vacuum expectation value (VEV) for $H_{2_3}$ is
taken with the alignment

\begin{equation}
<H_{2_3}> = V_{2_3} (1, 1)
\label{vev}
\end{equation}
while as in \cite{FKM} the other VEVs include
\begin{equation}
<H_3> = V(1, -2, 1)
\label{H3vev}
\end{equation}

\bigskip

\section{Predictions of NMRT$^{'}$M (D)}

\bigskip

From the Yukawa term {\bf D} and the vacuum alignment we
can derive for the down-quark mass matrix

\begin{equation}
D = \left( \begin{array}{ccc}
M_b & \frac{1}{\sqrt{2}} Y_{{\cal Q}{\cal S}} V_{2_3} &   
\frac{1}{\sqrt{2}} Y_{{\cal Q}{\cal S}} V_{2_3}  \\  
0 &   
\frac{1}{\sqrt{3}} Y_{{\cal S}} V & -2 \sqrt{\frac{2}{3}} \omega Y_{{\cal S}} V \\
0 
&
\sqrt{\frac{2}{3}} Y_{{\cal S}} V &  \frac{1}{\sqrt{3}} \omega Y_{{\cal S}} V
\end{array}
\right)
\label{D}
\end{equation}
where $M_b=Y_bV_{1_3}$ and $\omega = e^{i\pi/3}$.

\bigskip

The hermitian squared mass matrix ${\cal D} \equiv D D^{\dagger}$ for the charge
$(-1/3)$ quarks is then

\begin{equation}
{\cal D} =
\left( \begin{array}{ccc}
M^{'2}_b & \frac{1}{\sqrt{6}} Y_{{\cal S}} Y_{{\cal Q}{\cal S}} V V_{2_3} (1-2\sqrt{2}\omega^2) &   
\frac{1}{\sqrt{6}} Y_{{\cal S}} Y_{{\cal Q}{\cal S}} V V_{2_3} (\omega^2 + \sqrt{2}) \\  
\frac{1}{\sqrt{6}} Y_{{\cal S}} Y_{{\cal Q}{\cal S}} V V_{2_3} (1 - 2\sqrt{2}\omega^{-2}) &   
3 (Y_{{\cal S}} V)^2 & -\frac{\sqrt{2}}{3} (Y_{{\cal S}} V)^2 \\
\frac{1}{\sqrt{6}} Y_{{\cal S}} Y_{{\cal Q}{\cal S}} V V_{2_3} (\omega^{-2} + \sqrt{2})   
&
-\frac{\sqrt{2}}{3} (Y_{{\cal S}} V)^2 &  (Y_{{\cal S}} V)^2 
\end{array}
\right)
\label{calD}
\end{equation}
where $M^{'2}_b = M_b^2 + (Y_{{\cal Q}{\cal S}} V_{2_3})^2$.

\bigskip

\noindent Note that in this model the mass matrix for the charge $+2/3$ quarks is diagonal
\footnote{This uses the approximation that the electron mass is $m_e=0$;
{\it c.f.} ref.\cite{FKM}.}
so the CKM mixing matrix arises purely from diagonalization of ${\cal D}$ in Eq.(\ref{calD}).
The presence of the complex $T^{'}$ Clebsch-Gordan
in Eq. (\ref{calD}) permits a Chen-Mahanthappa origin \cite{Chen}
for the
KM CP violating phase.

\bigskip

\noindent In Eq.(\ref{calD}) the $2\times2$ sub-matrix for the first two families
coincides with the result discussed earlier \cite{FKM} and hence the successful
Cabibbo angle formula $\tan 2\Theta_{12} = (\sqrt{3})/2$ is preserved as follows.

\bigskip

\noindent The relevant $2\times2$ submatrix of ${\cal D}$ is proportional to

\begin{equation}
{\cal D}_{2\times2}  =
\left( \begin{array}{cc}
3  & -\frac{\sqrt{2}}{3}  \\
-\frac{\sqrt{2}}{3}  &  1
\end{array}
\right)
\label{calD22}
\end{equation}

\bigskip

\noindent whose diagonalization leads to the Cabibbo angle formula

\begin{equation}
\tan 2\Theta_{12} = \sqrt{3}/2.
\label{Cabibbo}
\end{equation}

\bigskip

\noindent For $m_b^2$ the experimental value is $17.6 GeV^2$ \cite{PDG2008}
although the CKM angles and phase do not depend on this
overall normalization.

\bigskip

\noindent Actually our results depend only on assuming that the ratio
$(Y_{{\cal Q}{\cal S}} V_{2_3}/ Y_{{\cal S}} V)$ is much smaller than one.

\bigskip

Defining

\bigskip

\begin{equation}
{\cal D}^{'} = 3 {\cal D}/(Y_{\cal S} V)^2
\label{calDprime}
\end{equation}

\bigskip

we find

\bigskip

\begin{equation}
{\cal D}^{'}=
\left( \begin{array}{ccc}
{\cal D}^{'}_{11} & A e^{i\psi_1} & A \eta e^{i\psi_2} \\
A e^{-i\psi_1} & 9 & -\sqrt{2} \\
A \eta e^{-i\psi_2} &  -\sqrt{2} & 3
\end{array}
\right)
\label{Dprime}
\end{equation}
in which we denoted

\bigskip

\begin{equation}
{\cal D}^{'}_{11} = 3 M_b^{'2} /(Y_{\cal S}V)^2
\label{calD11}
\end{equation}

\bigskip

\begin{equation}
A = \left( \sqrt{\frac{3}{2}} \right) \left( \frac{Y_{\cal Q \cal S} V_{2_3}}{Y_{\cal S} V} \right)
|1-2 \sqrt{2} \omega^2|
\label{A}
\end{equation}

\bigskip

\begin{equation}
\eta = \left| \frac{\omega^2 + \sqrt{2}}{1-2\sqrt{2}\omega^2} \right| = 0.33615...
\label{eta}
\end{equation}

\bigskip

\begin{equation}
\tan \psi_1 = \frac{-\sqrt{6}}{1+\sqrt{2}}
= - 1.01461...
\label{psi1}
\end{equation}

\bigskip

\begin{equation}
\tan \psi_2 = \frac{\sqrt{3}}{2\sqrt{2}-1}
 = 0.94729...
 \label{psi2}
\end{equation}

\bigskip

\noindent To arrive at predictions for the other CKM mixing elements
other than the Cabibbo angle ({\it i.e.} $\Theta_{13},\Theta_{23},
\delta_{KM}$) one needs only to diagonalize the matrix ${\cal D}^{'}$
in Eq.(\ref{Dprime}) by 

\begin{equation}
{\cal D}^{'}_{diagonal} = V_{CKM}^{\dagger} {\cal D}^{'} V_{CKM}
\label{diagonal}
\end{equation}

\bigskip

\noindent We write the mixing matrix as

\begin{equation}
V_{CKM} = \left( \begin{array}{ccc}
1 & V_{ts} & V_{td} \\
V_{cb} & \cos\Theta_{12} & \sin\Theta_{12} \\
V_{ub} & -\sin\Theta_{12} & \cos\Theta_{12} 
\end{array}
\right)
\label{CKM}
\end{equation}

\bigskip

\noindent and substituting Eq.(\ref{CKM}) into
Eq.(\ref{diagonal}) and using Eq.(\ref{Dprime})
leads to

\begin{equation}
\left( \begin{array}{c}
V_{cb} \\
V_{ub}
\end{array} \right)
= \frac{1}{\hat{{\cal D}}_{11}^{'}}
\left( \begin{array}{cc} 
{\cal D}^{'}_{11} - 3 & -\sqrt{2} \\
-\sqrt{2} & {\cal D}^{'}_{11} - 9
\end{array}
\right)
\left( \begin{array}{c}
A e^{-i\psi_1} \\
A e^{-i\psi_2}
\end{array} \right)
\label{VV1}
\end{equation}
where $\hat{{\cal D}}_{11}^{'}
= ({\cal D}^{'}_{11} -6 - \sqrt{11})({\cal D}^{'}_{11} - 6 + \sqrt{11})$.

\bigskip

\noindent while from unitarity it follows that

\begin{equation}
\left( \begin{array}{c}
V_{ts} \\
V_{td}
\end{array} \right)
= -
\left( \begin{array}{cc}
\cos\Theta_{12} & - \sin\Theta_{12} \\
\sin\Theta_{12} & \cos\Theta_{12} 
\end{array}
\right)
\left( \begin{array}{c}
V_{cb}^{*} \\
V_{ub}^{*}
\end{array} \right)
\label{VV2}
\end{equation}

\bigskip

\noindent The strategy now is to calculate the CP-violating
Kobayashi-Maskawa phase given by

\begin{equation}
\delta_{KM} = \gamma =  \arg \left( - \frac{V_{ud}V_{ub}^{*}}{V_{cd}V_{cb}^{*}}
\right)
\label{KM}
\end{equation}

\bigskip

\noindent and using Eqs.(\ref{CKM},\ref{VV1}) we arrive at the formula
in terms of ${\cal D}_{11}$

\begin{equation}
\delta_{KM} = \gamma_{T^{'}}
= \arg \left[ \frac{-\sqrt{2} + ({\cal D}^{'}_{11}-9) \eta e^{-i(\psi_1 - \psi_2)}}
{({\cal D}^{'}_{11} - 3) - \sqrt{2} \eta e^{-i(\psi_1 -\psi_2)}} \right]
=
\arg [\Gamma({\cal D}_{11}^{'})]
\label{gammaTprime}
\end{equation}
where $\Gamma$, a function of ${\cal D}_{11}^{'}$, is defined for later use.

\bigskip

\noindent In Fig. 1, we show a plot of $\gamma_{T^{'}}$ versus ${\cal D}_{11}^{'}$
using Eq.(\ref{gammaTprime})
and taking the range of experimentally-allowed $\gamma \equiv \delta_{KM}$
from the global fit\cite{CKMfitter} prompts us to use a value
${\cal D}_{11}^{'} = 19 \pm 2$ in the subsequent analysis.

\bigskip

\bigskip

\begin{figure}
\begin{center}
\vspace{15pt}
\includegraphics[height=70mm]{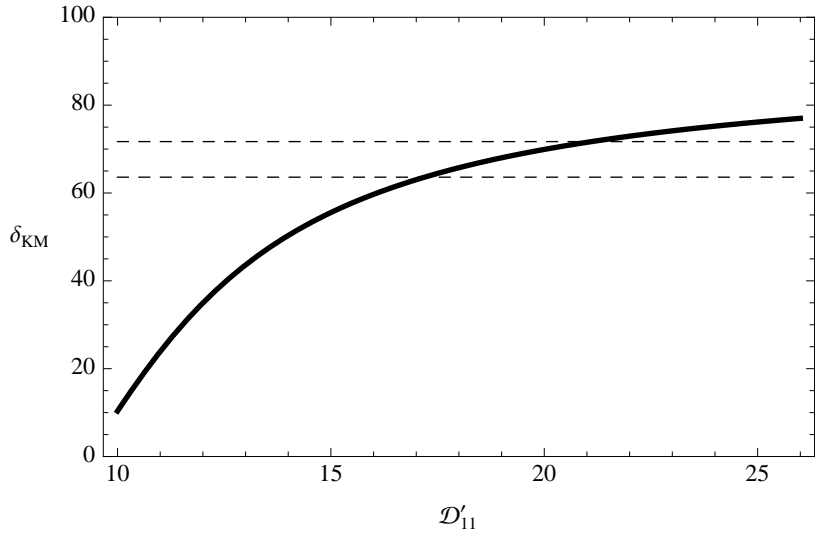}
\vspace{15pt}
\caption{\normalsize
The vertical axis is the value
of $\delta_{KM} \equiv \gamma_{T^{'}}$ in degrees and
the 
horizontal axis is the value of ${\cal D}^{'}_{11}$
defined in the text.
The dashed horizontal lines give the
$1\sigma$ range for $\delta_{KM}$ allowed
by the global fit of \cite{CKMfitter}.}
\label{FigureDprime}
\end{center}
\end{figure}

\bigskip

\noindent From the preceding equations
(\ref{CKM},\ref{VV1}) we find a formula for

\begin{equation}
|V_{ub}/V_{cb}| = |\tan\Theta_{13}\sin\Theta_{23}| 
\end{equation}

\bigskip

\noindent using unitarity, Eq.(\ref{VV2}), from the form
for the ratios of CKM matrix elements

\begin{equation}
|V_{td}/V_{ts}| = \left| \frac{ \sin\Theta_{12} + \Gamma({\cal D}_{11}^{'}) \cos\Theta_{12} }
{\cos\Theta_{12} - \Gamma({\cal D}_{11}^{'}) \sin\Theta_{12}} \right|
\end{equation}

\bigskip 

\newpage

\noindent Fig. 2 shows a plot of $|V_{td}/V_{ts}|$
as a function of ${\cal D}_{11}^{'}$.
It requires a value of ${\cal D}_{11}^{'}$
of approximately 16 which is sufficiently close
to that in Fig. 1.

\bigskip
\bigskip

\noindent For the value of $|V_{ub}/V_{cb}|$
there is approximately a factor two
between the prediction (higher)
and the best value from \cite{CKMfitter}.

\newpage

\begin{figure}
\begin{center}
\vspace{15pt}
\includegraphics[height=70mm]{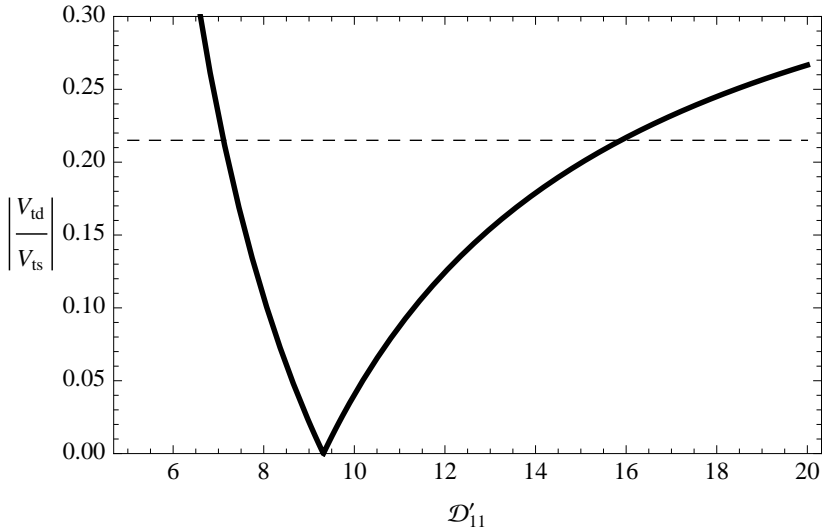}
\vspace{15pt}
\caption{\normalsize
The vertical axis is the value
of $|V_{td}/V_{ts}|$ and
the 
horizontal axis is the value of ${\cal D}^{'}_{11}$
defined in the text.
The dashed horizontal lines give the
value with small error allowed
by the global fit of \cite{CKMfitter}.}
\label{FigureVV}
\end{center}
\end{figure}

\bigskip

\newpage

\bigskip

\section{Discussion}

\bigskip

\noindent Note that once the off-diagonal
third-family elements
in Eq.(\ref{calD}) are taken as much smaller
than the elements involved in the
Cabibbo angle, the two KM angles and the CP phase
are predicted by the present NMRT$^{'}$M
in general agreement
so this vindicates the hope expressed in \cite{FKM}.

\bigskip

\noindent With regard to alternative NMRT$^{'}$M models discussed
earlier the possibilities {\bf A} and {\bf C} modify
the charge-2/3 mass matrix where we take flavor and
mass eigenstates coincident. The final possibility {\bf C}
does modify the charge (-1/3) mass matrix but does
not permit CP violation to arise from the Chen-Mahanthappa
mechanism as in the {\bf D} model
we have analysed both here and in \cite{FM1}.

\bigskip

\noindent With respect to the article \cite{FM1} which 
was letter length, the
present article presents more technical detail and figures
to clarify the results and predictions merely stated
in \cite{FM1} without explanation.

\bigskip

\noindent In summary, we have reported results of studying 
mixing angles by exploying the binary tetrahedral group ($T^{'}$)
as a global discrete flavor symmetry commuting with the 
local gauge symmetry $SU(3) \times SU(2) \times U(1)$ of
the standard model of particle phenomenology. The results
are encouraging to pursue this direction of study.

\newpage

\begin{center}

\section*{Acknowledgements}

\end{center}

This work was supported in part 
by the U.S. Department of Energy under Grant
No. DE-FG02-06ER41418.

\bigskip
\bigskip
\bigskip
\bigskip

\end{document}